\documentclass[sigconf]{acmart}

\AtBeginDocument{%
  \providecommand\BibTeX{{%
    \normalfont B\kern-0.5em{\scshape i\kern-0.25em b}\kern-0.8em\TeX}}}

\usepackage{booktabs} 
\usepackage[utf8]{inputenc}
\usepackage[russian,english]{babel}

\setcopyright{rightsretained}
\copyrightyear{2019}
\acmYear{2019}
\acmDOI{}

\setcopyright{rightsretained}
\copyrightyear{2020}
\acmYear{2020}
\acmConference{SIGGRAPH '20 Emerging Technologies}{August 17, 2020}{Virtual Event, USA}\acmBooktitle{Special Interest Group on Computer Graphics and Interactive Techniques Conference Emerging Technologies (SIGGRAPH '20 Emerging Technologies), August 17, 2020}
\acmISBN{978-1-4503-7967-0/20/08}
 
\begin{document}
\title{TeslaMirror: Multistimulus Encounter-Type Haptic Display for Shape and Texture Rendering in VR}
\author{Aleksey Fedoseev}
\affiliation{%
  \institution{Skolkovo Institute of Science and Technology (Skoltech)}
  \streetaddress{Nobelya Ulitsa 3}
  \city{Moscow}
  \country{Russia}
  \postcode{121205}
}
\email{aleksey.fedoseev@skoltech.ru}

\author{Akerke Tleugazy}
\affiliation{%
  \institution{Skolkovo Institute of Science and Technology (Skoltech)}
  \streetaddress{Nobelya Ulitsa 3}
  \city{Moscow}
  \country{Russia}
  \postcode{121205}
}
\email{akerke.tleugazy@skoltech.ru}

\author{Luiza Labazanova}
\affiliation{%
  \institution{Hong Kong Polytechnic University (PolyU)}
  \streetaddress{Hung Hom}
  \city{Kowloon}
  \country{Hong Kong}
  \postcode{}
}
\email{luiza.labazanova@connect.polyu.hk}

\author{Dzmitry Tsetserukou}
\affiliation{%
  \institution{Skolkovo Institute of Science and Technology (Skoltech)}
  \streetaddress{Nobelya Ulitsa 3}
  \city{Moscow}
  \country{Russia}
  \postcode{121205}
}
\email{d.tsetserukou@skoltech.ru}

\renewcommand{\shortauthors}{Fedoseev, Tleugazy, Labazanova and Tsetserukou}

\begin{abstract}
This paper proposes a novel concept of a hybrid tactile display with multistimulus feedback, allowing the real-time experience of the position, shape, and texture of the virtual object. The key technology of the TeslaMirror is that we can deliver the sensation of object parameters (pressure, vibration, and shear force feedback) without any wearable haptic devices. We developed the full digital twin of the 6 DOF UR robot in the virtual reality (VR) environment, allowing the adaptive surface simulation and control of the hybrid display in real-time. 

The preliminary user study was conducted to evaluate the ability of TeslaMirror to reproduce shape sensations with the under-actuated end-effector. The results revealed that potentially this approach can be used in the virtual systems for rendering versatile VR shapes with high fidelity haptic experience.

\end{abstract}



\begin{teaserfigure}
  \includegraphics[width=\textwidth]{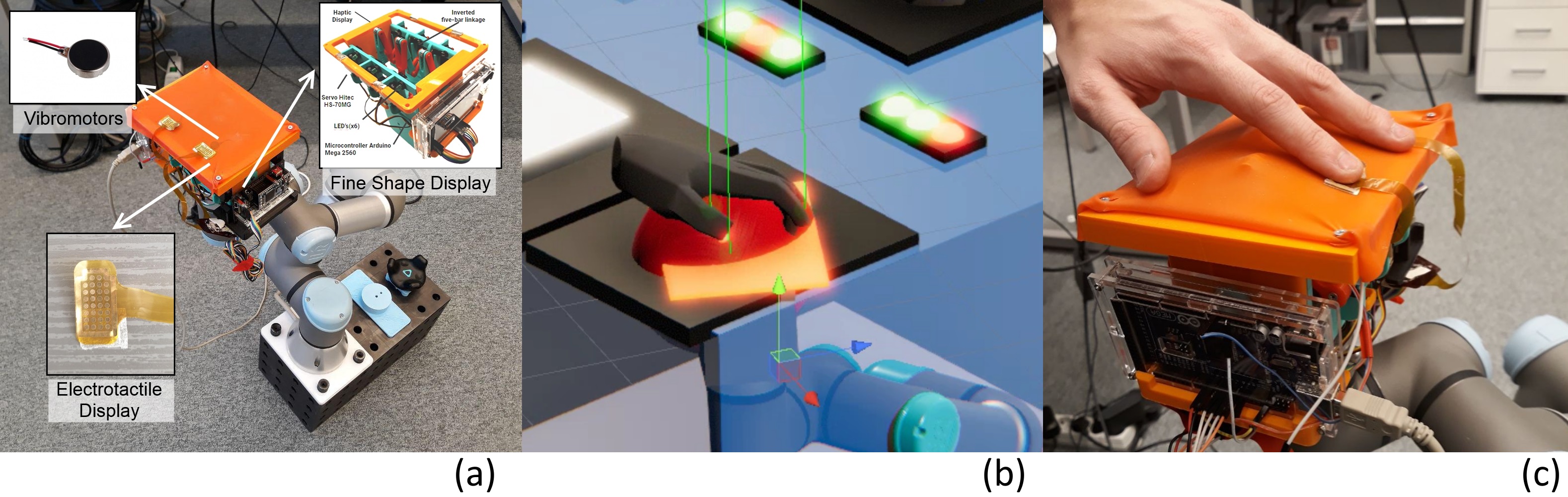}
  \caption{(a) TeslaMirror architecture. (b) User approaching the virtual surface. (c) Real-time simulation by haptic display.}
  \label{fig:ill5}
\end{teaserfigure}

\maketitle
\section{Introduction}
The ability to touch and feel virtual objects is essential in the training and entertainment industry, therefore, the importance of realistic object sensation is extremely high. The solution with a robotic display was proposed by Bruno Araujo et al., the design of Snake Charmer encounter-type display allows the realistic presentation of the positional and texture feedback of the VR surface  \cite{Araujo:2016:SCP:2839462.2839484}. However, the requirements for hardware replacement of the texture sample makes it impossible to use such display for continuous contact with any complex shape. 
For more realistic rendering, researchers reproduced vibration and friction that occur when interacting with objects. They developed devices based on the stylus controller, touchscreen, or wearable haptic glove to sense objects in the virtual environment \cite{10.1145/1866029.1866074}. 
Although existing solutions showed good results in textures recognition and perceptual similarity, they require the use of additional tools.

\section{Mechanical impact: gross roughness and position feedback}
We designed a proof-of-concept prototype based on the combination of a TeslaMirror system and a haptic texture sensation device. The developed MirrorShape framework consists of a 6 DOF Universal robot UR3, PC with control framework, tracking system, and shape-forming end-effector as haptic display \cite{10.1145/3359996.3365049}. When the user's hand approaches an object in VR the framework estimates the position and orientation of the end-effector as normal to the palm orientation (Figure 2).

\begin{figure}[h!]
\centering
  \includegraphics[width=0.8\columnwidth]{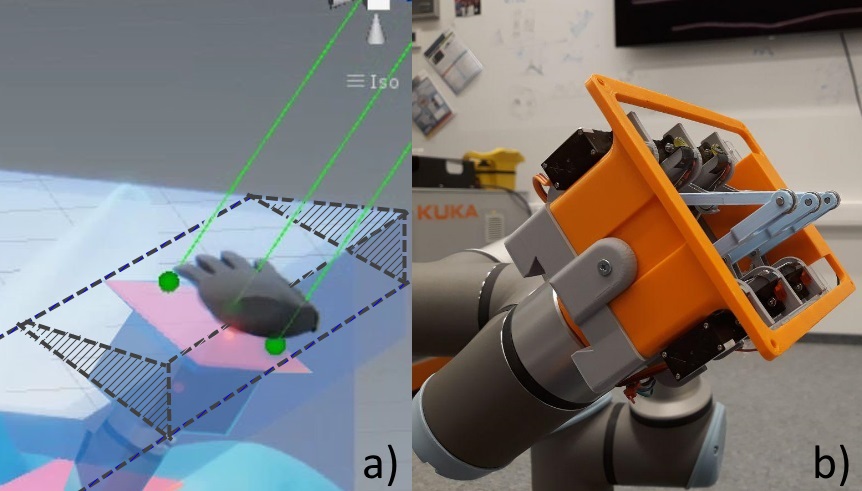}
  \caption{(a) Calculation of the collision point, (b) TeslaMirror performing surface simulation}~\label{fig:figure4}
\end{figure}

The design of an inverted five-bar linkage mechanism allows TeslaMirror to define the position of 3 contact points independently and combined with a flexible display that provides interaction with various shapes only with 3 DOF devices instead of using a large number of pins and actuators.

\section{Vibrotactile and Electrotactile impact: texture roughness and direction feedback}
 Our texture display device is based on the electro-vibration system of Yem and Kajimoto \cite{inproceedings} that merge mechanical and electrical stimulation of skin mechanoreceptors. The electrical stimulation is used to provide low-frequency vibration and sensation of direction. The device consists of a 4x5 electrode array film and controller board. Vibration motors were used to simulate high-frequency vibration and skin deformation. We have attached two ERM vibrotactile actuators and electrode array to the board.

\section{User Study}

To evaluate the performance under-actuated design developed for the fine shape display we investigated the user's perception of different virtual shapes simulated by 3 contact points of TeslaMirror. For this study 7 right-handed subjects was chosen with the age 22 to 30. Four different shapes were chosen: sphere, cube, pyramid, and edge. Each of them was rendered by the VR framework and evaluated by the subject total 20 times in random order (Figure 3). 

\begin{figure} [h]
  \centering
  \includegraphics[width=0.75\linewidth]{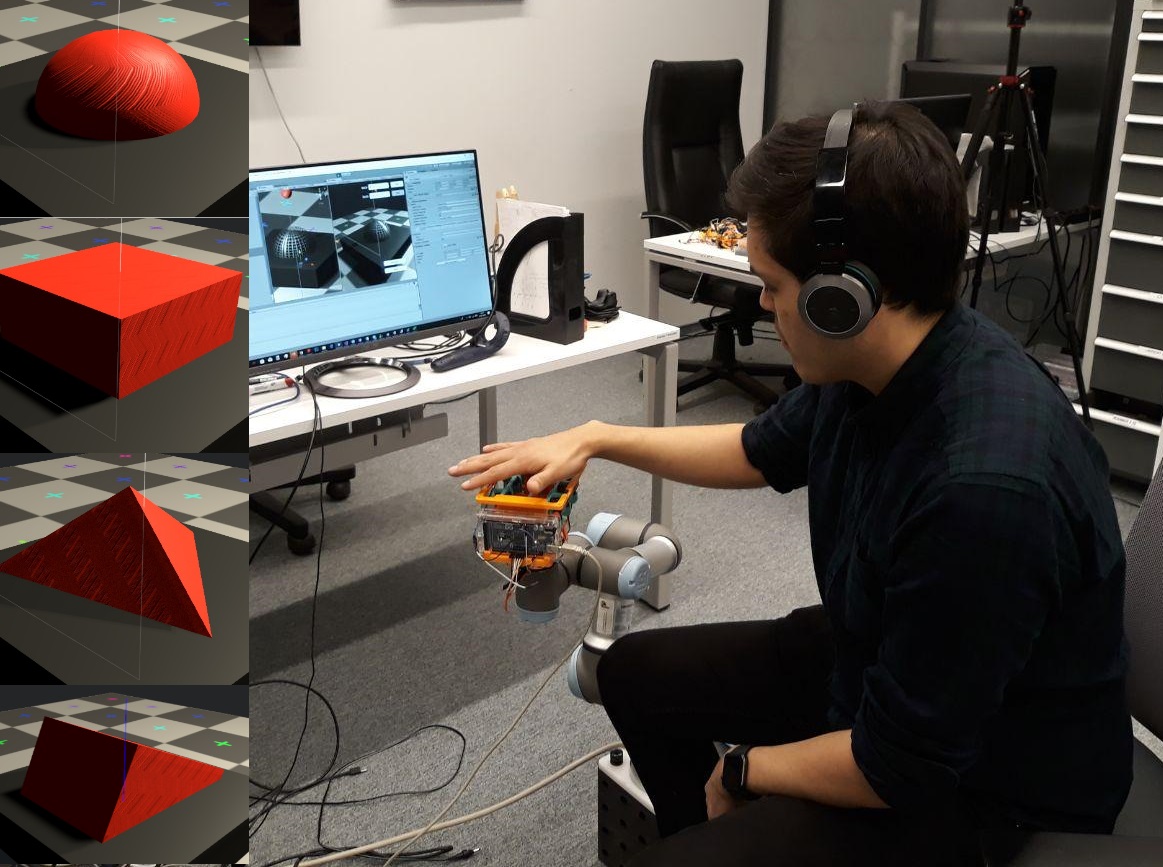}
  \caption{ User study setup. }
  \label{fig:patterns}
\end{figure}

The ANOVA results of the evaluation (Table 1) showed a statistically significant difference in recognition rates for different rendered shapes $(F(3,21) = 8.11, p = 7.48\cdot10^{-4}<0.05)$. The least recognizable was the difference between spherical and pyramid shapes.

\begin{table}[h]
    \centering
    \caption{Confusion matrix of different virtual shapes.}
    \includegraphics[width=0.9\linewidth]{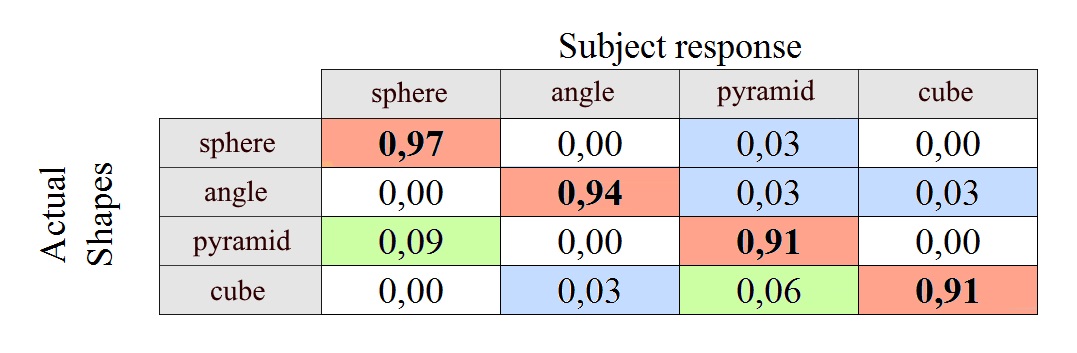}
    \label{tab:my_label}
\end{table}
\section{Acknowledgements} 
\begin{otherlanguage*}{russian}
The authors thank Professor Hiroyuki Kajimoto and Professor Vibol Yem for developing and providing the electrotactile device. The reported study was funded by FASIE, project number №606ГУЦЭС8-D3/62112.
\end{otherlanguage*}
\bibliographystyle{ACM-Reference-Format}
\bibliography{sample-bibliography}


\begin{thebibliography}{4}


\ifx \showCODEN    \undefined \def \showCODEN     #1{\unskip}     \fi
\ifx \showDOI      \undefined \def \showDOI       #1{#1}\fi
\ifx \showISBNx    \undefined \def \showISBNx     #1{\unskip}     \fi
\ifx \showISBNxiii \undefined \def \showISBNxiii  #1{\unskip}     \fi
\ifx \showISSN     \undefined \def \showISSN      #1{\unskip}     \fi
\ifx \showLCCN     \undefined \def \showLCCN      #1{\unskip}     \fi
\ifx \shownote     \undefined \def \shownote      #1{#1}          \fi
\ifx \showarticletitle \undefined \def \showarticletitle #1{#1}   \fi
\ifx \showURL      \undefined \def \showURL       {\relax}        \fi
\providecommand\bibfield[2]{#2}
\providecommand\bibinfo[2]{#2}
\providecommand\natexlab[1]{#1}
\providecommand\showeprint[2][]{arXiv:#2}

\bibitem[\protect\citeauthoryear{Araujo, Jota, Perumal, Yao, Singh, and
  Wigdor}{Araujo et~al\mbox{.}}{2016}]%
        {Araujo:2016:SCP:2839462.2839484}
\bibfield{author}{\bibinfo{person}{Bruno Araujo}, \bibinfo{person}{Ricardo
  Jota}, \bibinfo{person}{Varun Perumal}, \bibinfo{person}{Jia~Xian Yao},
  \bibinfo{person}{Karan Singh}, {and} \bibinfo{person}{Daniel Wigdor}.}
  \bibinfo{year}{2016}\natexlab{}.
\newblock \showarticletitle{Snake Charmer: Physically Enabling Virtual
  Objects}. In \bibinfo{booktitle}{\emph{Proceedings of the TEI '16: Tenth
  International Conference on Tangible, Embedded, and Embodied Interaction}}
  \emph{(\bibinfo{series}{TEI '16})}. \bibinfo{publisher}{ACM},
  \bibinfo{address}{New York, NY, USA}, \bibinfo{pages}{218--226}.
\newblock
\showISBNx{978-1-4503-3582-9}
\urldef\tempurl%
\url{https://doi.org/10.1145/2839462.2839484}
\showDOI{\tempurl}


\bibitem[\protect\citeauthoryear{Bau, Poupyrev, Israr, and Harrison}{Bau
  et~al\mbox{.}}{2010}]%
        {10.1145/1866029.1866074}
\bibfield{author}{\bibinfo{person}{Olivier Bau}, \bibinfo{person}{Ivan
  Poupyrev}, \bibinfo{person}{Ali Israr}, {and} \bibinfo{person}{Chris
  Harrison}.} \bibinfo{year}{2010}\natexlab{}.
\newblock \showarticletitle{TeslaTouch: Electrovibration for Touch Surfaces}.
  In \bibinfo{booktitle}{\emph{Proceedings of the 23nd Annual ACM Symposium on
  User Interface Software and Technology}} \emph{(\bibinfo{series}{UIST
  ’10})}. \bibinfo{publisher}{Association for Computing Machinery},
  \bibinfo{address}{New York, NY, USA}, \bibinfo{pages}{283–292}.
\newblock
\showISBNx{9781450302715}
\urldef\tempurl%
\url{https://doi.org/10.1145/1866029.1866074}
\showDOI{\tempurl}


\bibitem[\protect\citeauthoryear{Fedoseev, Chernyadev, and
  Tsetserukou}{Fedoseev et~al\mbox{.}}{2019}]%
        {10.1145/3359996.3365049}
\bibfield{author}{\bibinfo{person}{Aleksey Fedoseev}, \bibinfo{person}{Nikita
  Chernyadev}, {and} \bibinfo{person}{Dzmitry Tsetserukou}.}
  \bibinfo{year}{2019}\natexlab{}.
\newblock \showarticletitle{Development of MirrorShape: High Fidelity
  Large-Scale Shape Rendering Framework for Virtual Reality}. In
  \bibinfo{booktitle}{\emph{25th ACM Symposium on Virtual Reality Software and
  Technology}} \emph{(\bibinfo{series}{VRST ’19})}.
  \bibinfo{publisher}{Association for Computing Machinery},
  \bibinfo{address}{New York, NY, USA}, Article \bibinfo{articleno}{Article
  105}, \bibinfo{numpages}{2}~pages.
\newblock
\showISBNx{9781450370011}
\urldef\tempurl%
\url{https://doi.org/10.1145/3359996.3365049}
\showDOI{\tempurl}


\bibitem[\protect\citeauthoryear{Yem and Kajimoto}{Yem and Kajimoto}{2017}]%
        {inproceedings}
\bibfield{author}{\bibinfo{person}{Vibol Yem} {and} \bibinfo{person}{Hiroyuki
  Kajimoto}.} \bibinfo{year}{2017}\natexlab{}.
\newblock \showarticletitle{Wearable tactile device using mechanical and
  electrical stimulation for fingertip interaction with virtual world}.
  \bibinfo{pages}{99--104}.
\newblock
\urldef\tempurl%
\url{https://doi.org/10.1109/VR.2017.7892236}
\showDOI{\tempurl}


\end{thebibliography}

\end{document}